\documentclass[preprint,showpacs,floats,letterpaper,floatfix,,groupedaddress,eqsecnum]{revtex4}
\usepackage{amssymb,amsmath}
\usepackage[dvips]{graphicx}

\usepackage{dcolumn,epsfig}

\begin{document}

\title{ Ratchet effect in inhomogeneous inertial systems: II. The square-wave drive case.}

\author{S. Saikia$^1$ and Mangal C. Mahato}
\email{mangal@nehu.ac.in}
\affiliation{Department of Physics, North-Eastern Hill University, Shillong-793022,
India\\
and\\
$^1$Department of Physics, St. Anthony's College, Shillong-793001, India.}

\begin{abstract}
The underdamped Langevin equation of motion of a particle, in a symmetric periodic 
potential and subjected to a symmetric periodic forcing with mean zero over a period, 
with nonuniform friction, is solved numerically. The particle is shown to acquire a 
steady state mean velocity at asymptotically large time scales. This net particle 
velocity or the ratchet current is obtained in a range of forcing amplitudes $F_0$ and 
peaks at some value of $F_0$ within the range depending on the value of the average 
friction coefficient and temperature of the medium. At these large time scales the 
position dispersion grows proportionally with time, $t$, allowing for calculating the 
steady state diffusion coefficient $D$ which, interestingly, shows a peaking behaviour 
around the same $F_0$. The ratchet current, however, turns out to be largely coherent. 
At intermediate time scales, which bridge the small time scale behaviour of 
dispersion$\sim t^2$ to the large time one, the system shows, in some cases,  periodic 
oscillation between dispersionless and steeply growing dispersion depending on the 
frequency of the forcing. The contribution of these different dispersion regimes to 
ratchet current is analysed.

\end{abstract}

\vspace{0.5cm}
\date{\today}

\maketitle

\section{Introduction}
The investigation of particle motion in periodic potentials has obvious relevance in 
condensed matter studies. Motion of ions in a crystalline lattice is a case in point. 
Stochasticity in the motion is naturally introduced at nonzero temperatures. In these
environment the particle motion can be approximately described by a Langevin equation 
with suitable model potentials. Depending on the problem at hand the motion is either 
considered heavily damped, almost undamped, or in the intermediate situation mildly
damped (or underdamped). In many a situations in the former two extreme cases the 
Langevin equation becomes amenable to analytical solution. However, in the underdamped 
situation, barring a few special cases, numerical methods are used to solve the equation 
of motion of the particle\cite{risken}. Owing to various kinds of errors and 
approximations involved in these (numerical) methods exact quantitative solutions are 
not possible. However, the method can reveal useful qualitative trends in the behaviour 
of the particle motion. For instance, recently it was shown\cite{machura} that a 
particle, moving in an asymmetric but periodic potential in the presence of thermal 
noise when subjected to a symmetric periodic external drive (which adds to zero when 
averaged over a period), acquires a net motion when the parameters of the problem are 
chosen suitably. Such a net particle current without the application of any external 
bias or potential gradient in the presence of thermal noise is called thermal ratchet 
current and the system giving such a current is termed as thermal ratchet\cite{reim}. 
Here the equilibrium condition of detailed balance is not applicable because the system 
was driven far away from equilibrium by rocking it periodically in the presence of noise.
It was further shown that this system can even exhibit absolute negative 
mobility\cite{kostur}. This prediction has already been found to be true 
experimentally\cite{nagel}. It shows that in underdamped conditions or in the inertial 
regime diverse possibilities can be (qualitatively) uncovered by (numerically) solving 
the equations of motion.

In the above important example the potential asymmetry was one of the necessary 
conditions for realization of ratchet current. The particle had to surmount the same 
potential barrier on either direction; only the slopes leading to the top of the barrier
differed. A sinusoidal potential, for example, having no such asymmetry would not have 
yielded the ratchet current. In the present work, we consider similar particle motion in 
a sinusoidal potential. However, instead of a uniform friction coefficient of the medium 
we consider a model nonuniform space-dependent friction coefficient $\gamma (x)$ of the 
medium. In particular, we consider a sinusoidally varying $\gamma (x)$ exactly similar 
to the potential but with a phase lag, $\phi$. A simple illustrative example of the 
model can be imagined thus: a stationary pressure wave is established in air giving a 
periodic $\gamma (x)$ for particle motion along $x$. An array of ions with the 
periodicity of $\gamma(x)$ but shifted a little to give a phase lag $\phi$ will just fit
our model for a charged particle motion along $x$.  Here the potential is symmetric and 
periodic. However, the directional symmetry of the system is disturbed by a phase shift 
in the similarly periodic $\gamma (x)$. Particle motion along an one dimensional 
semiconductor heterostructure or protein motor motion on the surface of a microtubule 
along its axis would be some practical situations close to the above example. Though we 
do not have any microscopic basis for justifying the model form of $\gamma (x)$ a 
periodic variation of friction has been argued earlier from mode-coupling theory of 
adatom motion on the surface of a crystal of identical atoms\cite{wahn}. Also, the 
equation of motion has a direct correspondence with the resistively and capacitatively 
shunted junction (RCSJ) model of Josephson junctions; the term describing the 
nonuniformity of friction having an one-to-one correspondence with the '$\cos\phi$' term 
in the RCSJ model\cite{falco}. A qualitative physical argument for the possibility of 
obtaining ratchet current in this inhomogeneous system was given in an earlier 
work\cite{wanda}.

We drive the system with a square-wave periodic field. The resulting Langevin equation
is solved numerically. We obtain
particle current and properties associated with it in the parameter space of external
field amplitude $F_0$, the average friction coefficient $\gamma_0$, the phase lag $\phi$,
and the temperature $T$. Since it is a formidable task to explore the entire parameter
space, we present results for only some regions of a few sections of this space where
appreciable ratchet current is obtained.

The ratchet current is obtained in the steady state situation which is achieved in
the asymptotic time limit. In our case we observe particle motion for a long time
$t$ such that the position dispersion $\langle(\Delta x(t))^2\rangle$ averaged over many
similar trajectories reach the situation where $\langle(\Delta x(t))^2\rangle \sim t$.
If the phase lag $\phi$ is considered equal to $n\pi$ $(n=0, 1, 2, ...)$ there will be
no ratchet current, $\bar v$, since in this case both the directions of $x$ are identical
in all respects. However, when $\phi \ne n\pi$ appreciable $\bar v$ is obtained
in a small range of $F_0$ with a peak in an intermediate $F_0$ for given $\gamma_0$,
$T$, and $\phi$.

In an earlier work\cite{wanda} the variation of ratchet current $\bar v$ as a function
of the amplitude $F_0$ of the applied square-wave forcing $F(t)$, with a frequency
of $5\times10^{-4}$ cycles per unit time,
was shown (Fig.5 of Ref.\cite{wanda}) for the two cases, adiabatic drive and square-wave
drive for $\gamma_0=0.035$ and temperature $T=0.4$. The range of $F_0$ over which
ratchet current is obtained in the square-wave drive case was wider [$0.1<F_0<0.8$] 
compared to the adiabatic drive condition [$0.07<F_0<0.12$] and the peak current also
occurs at a larger $F_0$ value. The range, though wider, still remains well below 
$F_c$, the critical field at which the potential barrier to motion just disappears. 
The current is, therefore, essentially aided by thermal noise.

Since in the steady state $\langle(\Delta x(t))^2\rangle \sim t$ we can define the
diffusion constant $D$: $\langle(\Delta x(t))^2\rangle =2Dt$. Interestingly, for given
$\gamma_0, T$, and $\phi$, the diffusion constant $D$ shows nonmonotonic behaviour with
the field amplitude $F_0$ and it peaks around a value of $F_0$ where $\bar v$ attains
maximum at the drive frequency of $5\times10^{-4}$ cycles per unit time. That is to say 
the ratchet current is maximised when the system is most
diffusive. To compare the extent of this diffusive spread with the directional average
displacement a quantity, P\'{e}clet number $P_e$, is defined as the ratio of square of
mean displacement $\bar x$ in time $t$ to half the square of diffusive spread in the
same time interval $t$ \cite{lind}:
\begin{equation}
P_e=\frac{{\bar x}^2}{Dt}=\frac{{\bar x}{\bar v}}{D}.
\end{equation}
Thus, if $P_e> 2$ the motion is dominated by directional transport and hence it is 
considered coherent otherwise the net displacement is overwhelmed by diffusive motion. 
Our calculation shows that in the region where the ratchet current is appreciable and in
particular where $\bar v$ peaks $P_e$ is much larger than 2. Here the ratchet current is 
obtained when there was neither a bias to help the system nor any load to oppose it. The 
current, however, is not large enough and when  a small load is applied against current 
the current either reduces to a small level or starts flowing in the direction of the 
applied load. Thus, in the given circumstances, no appreciable useful work can be 
extracted from this inhomogeneous (frictional) ratchet. However, even in the absence of 
any external load the particle keeps moving against the frictional resistance. Leaving 
out the symmetric diffusive part of the motion the particle's unidirectional (ratchet) 
current $\bar v$ is maintained against the average frictional force. The ratio of this 
work (the ratchet performs against the frictional drag) to the total energy pumped into 
the system from the source of the external forcings is termed as {\it Stokes efficiency},
$\eta_{S}$, of the ratchet.

An expression for $\eta_{S}$ has been derived earlier\cite{machura, lind, seki}
which involves $\bar v$ as well as the second moment of the velocity $v$ calculated from
the probability distribution $P(v)$ of the velocity $v(t)$ recorded all through the
trajectory of the particle. We have calculated $\eta_{S}$ as a function of the
amplitude of the applied forcing. $\eta_{S}$ shows a peak, the position of which,
however, does not coincide with the peak of the ratchet current. The distribution $P(v)$
is almost symmetric about $v=0$ and the velocity dispersion grows monotonically
approaching to be linear in $F_0$ at large $F_0$.

Recently, it has been reported\cite{linde} that in a tilted periodic potential an
underdamped particle motion shows dispersionless behaviour in the intermediate time
regime for a range of tilt values. In the present work we show that when the system is
driven by a square-wave forcing of appropriate amplitude such dispersionless transient
behaviour with added richness can be observed. The dispersionless behaviour of constant
tilt gets punctuated and oscillatory behaviour of dispersion of different kinds,
depending on the frequency of the periodic drive, naturally emerges. Interestingly, 
however, contrary to expectations, dispersionless particle motion do not contribute to
(instead hinders) ratchet current in this system. 

In section II the basic equation of motion used in this model calculation will be
presented. The section III will be devoted to the presentation of the detailed results of
our numerical calculation. In the last section (Sec. IV) we shall conclude with a 
discussion.

\section{The model}

This part (II) of the work is an extension of our earlier work (part I), where the
system was driven adiabatically\cite{wanda1} to obtain ratchet current. In the present
case we drive the system periodically by a symmetric square-wave forcing and calculate
the particle trajectories $x(t)$. The choice of
square-wave forcing, instead of a sinusoidal forcing, is to make a direct contact with 
the adiabatically driven case. We thus have the forcing $F(t)$ as,\\
\[ \begin{array}{llll}
  F(t)&=&\pm F_0,& ( n\tau\leq t < (n+\frac{1}{2})\tau),\\
 &=&\mp F_0,& ( (n+\frac{1}{2})\tau \leq t < (n+1)\tau),\\
\end{array}\] \\
where $\tau$ is the period of forcing and $n= 0, 1, 2, ... $. In what follows we shall 
refer to trajectories $x(t)$ computed using the upper signs as the odd numbered 
trajectories and those calculated with the lower signs as the even numbered trajectories.

The motion of a particle of mass $m$ moving in a periodic potential $V(x)=-V_0 sin(kx)$
in a medium with friction coefficient $\gamma (x)=\gamma_0(1-\lambda sin(kx+\phi))$
with $0\leq \lambda<1$ and subjected to a square-wave forcing $F(t)$ is described by the
Langevin equation\cite{amj1},
\begin{equation}
m\frac{d^{2}x}{dt^{2}}=-\gamma (x)\frac{dx}{dt}-\frac{\partial{V(x)}}{\partial
x}+F(t)+\sqrt{\gamma(x)T}\xi(t).
\end{equation}
Here $T$ is the temperature in units of the Boltzmann constant $k_B$. The Gaussian 
distributed fluctuating forces $\xi (t)$ satisfy the statistics: $<\xi (t)>=0$, and 
$<\xi(t)\xi(t^{'})>=2\delta(t-t^{'})$. For convenience, we write down Eq.(2.1) in 
dimensionless units by setting $m=1$, $V_0=1$, $k=1$ so that $T=2$ corresponds to an 
energy equivalent equal to the potential barrier height at $F_0=0$. The reduced Langevin
equation, with reduced variables denoted again by the same symbols, is written now as
\begin{equation}
\frac{d^{2}x}{dt^{2}}=-\gamma(x)\frac{dx}{dt}
+cos x +F(t)+\sqrt{\gamma(x) T}\xi(t),
\end{equation}
where $\gamma(x)=\gamma_0(1-\lambda sin(x+\phi))$. Thus the periodicity of the potential 
$V(x)$ and also the friction coefficient $\gamma$ is $2\pi$ \cite{comp}. The potential 
barrier between any two consecutive wells of $V(x)$ persists for all $F_0<1$ and it just
disappears at the critical field value $F_0=F_c=1$. The noise variable, in the same 
symbol $\xi$, satisfies exactly similar statistics as earlier.

The Eq. (2.2) is solved numerically (with given initial conditions) to obtain the 
trajectory $x(t)$ of the particle for various values of the parameters $F_0, \gamma_0$, 
and $T$. Also, the steady state mean velocity $\bar v$ of the particle is obtained as
\begin{equation}
 {\bar v}=   \langle \lim_{t \rightarrow \infty}\frac{x(t)}{t}\rangle,
\end{equation}
where the average $\langle ... \rangle$ is evaluated over many trajectories. The mean
velocity is also calculated from the distribution $P(v)$ of velocities giving almost
identical result.

\section{Numerical Results}
The Langevin equation (2.2) is solved numerically using two methods: 4$^{th}$-order 
Runge-Kutta\cite{nume} and Heun's method (for solving ordinary differential equations). 
We take a time step interval of 0.001 during which the fluctuating force $\xi (t)$, 
obtained from a Gaussian distributed random number appropriate to the temperature $T$, 
is considered as 
constant and the equation solved as an initial value problem. In the next interval 
another random number is called to use as the value of $\xi$ and the process repeated. 
By a careful observation of the individual trajectories of the particle shows that by 
$t=10^4$ the particle completely loses its memory of the initial condition it had 
started with. When we look for steady state solutions the trajectory is generally 
allowed to run for a time $t\sim 10^7$. Therefore, for steady state evaluation for 
$\bar v$, etc. the results become independent of initial conditions. The (Runge-Kutta) 
method had earlier been used and obtained correct results\cite{borrom} in a similar 
situation. Also, the method was checked against results obtained earlier for the 
adiabatic case (using matrix continued fraction method) and found to compare well 
qualitatively\cite{wanda}. Heun's method when applied in similar situations take much 
less time than the Runge-Kutta method and yields qualitatively as good result (Fig.1). 
With this confidence in our numerical procedures, we apply either one or the other of 
these two numerical schemes as the situation demands. We take $\lambda=0.9$, 
$\gamma_0=0.035$, and $T=0.4$ all through our calculation in the following.

The motion of the particle is governed by the applied square-wave  forcing $F(t)$. As 
$F(t)$ changes periodically so does the position of the particle. In view of this effect
we start our simulation at $t=0$ with $F(0)=+|F_0|$, and $-|F_0|$ for alternate 
trajectories (to be referred to respectively as odd numbered and even numbered ones, as
mentioned earlier). This gives a nice nonoscillating variation of the overall average 
position when averaged over an large even number of trajectories. However, while 
calculating the position dispersions or velocity dispersions at a given time $t$, the 
even and odd trajectories are treated separately to calculate the deviations from their 
respective mean values.
\subsection{The ratchet current} 
\begin{figure}
\begin{center}
\epsfig{file=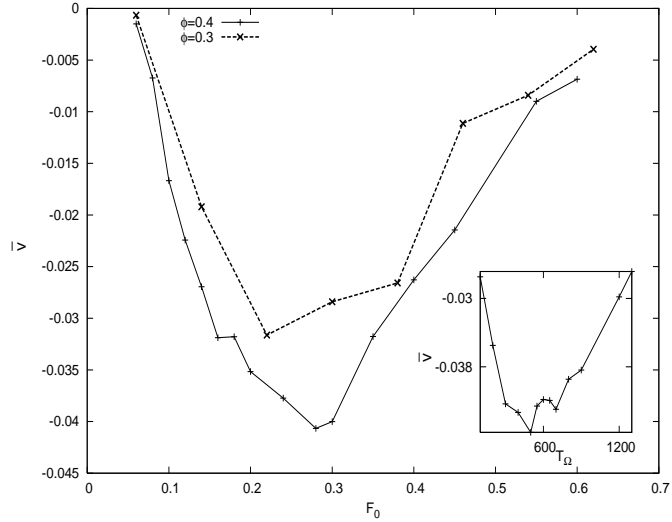, angle=-90, width=9cm,totalheight=7cm}
\caption{Shows the variation of $\bar v$ with $F_0$ for phase lag $\phi=0.3$ and 0.4 with
$T_{\Omega}=1000$.
The inset shows the variation of $\bar v$ with time period $T_{\Omega}$ for $F_0=0.26$.
\label{fig1}}
\end{center}
\end{figure}

In Fig.1, $\bar v(F_0)$ are plotted for two values of phase lag $\phi =0.3$, and 0.4 for 
the same values of $\gamma_0$ and $T$ to illustrate the effect of $\phi$ for 
$\tau = 2000$. The ratchet current $\bar v$ also shows nonmonotonic behaviour as 
a function of the period of the drive. In the inset (Fig. 1) we plot the varation of 
$\bar v$ as a function of the time period of the drive for $F_0=0.26$. For these 
parameter values the current $\bar v$ peaks at period $\tau \approx 1000$. For 
comparison of time scales, it may be noted that for an equivalent RCSJ model of 
Josephson junctions the characteristic Josephson plasma frequency $\omega_J$ turns out 
to be about $10^3$ times larger than the drive frequency corresponding to 
$T_{\Omega}=1000$, where $T_{\Omega}=\frac{\tau}{2}$. (The sign of $F_0$ is changed 
after every time interval $T_{\Omega}$.) In this sense we obtain appreciable ratchet 
current only for very slow drives. It should, however, be noted that in the infinitely 
slow adiabatic case the ratchet current is effectively zero for $F_0>0.12$ for 
$\gamma_0=0.035$ at $T=0.4$. In the following we shall present the results obtained 
using $\phi=0.35$ except when mentioned otherwise explicitly. 

\subsection{The steady-state dispersions}
The position dispersions $\langle (\Delta x(t))^2 \rangle$, where $\Delta x(t) =
x(t)-\langle x(t) \rangle$ are evaluated over a large number of trajectories for various
values of $F_0$, and $T_{\Omega}=1000$. It is found that the dispersions fit nicely to
\begin{equation}
\log[\langle (\Delta x(t))^2 \rangle]=\log(t)+\log(2D),
\end{equation}
for large $t$, typically $t>10^5$ (Fig. 6 of \cite{wanda}).

\begin{figure}
\begin{center}
\epsfig{file=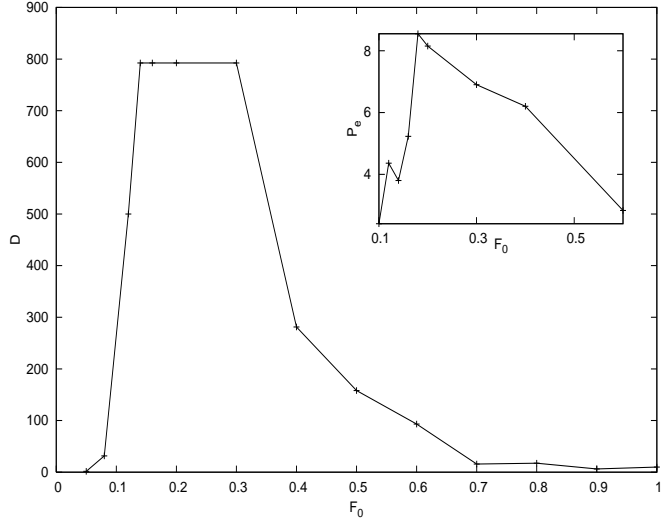, angle=-90, width=9cm,totalheight=7cm}

\caption{The variation of the diffussion constant D as a function of the driving 
amplitude $F_0$ with $T_{\Omega}=1000$. 
The inset shows the variation of the corresponding P\'{e}clet number $P_e$ with $F_0$.
\label{fig2}}

\end{center}
\end{figure}

 From the linear fit of the 
graphs we calculate the diffusion constants $D(F_0)$ and the result is shown in Fig.2. 
The diffusion constant has a large value between $F\approx 0.15$ and 0.35. The peak 
height is quite large $\approx 800$. As $F_0$ is increased $D$ decreases sharply and 
becomes smaller than 50 (which is less than 10\% of its peak value) for $F_0>0.7$. This 
$[0.15\leq F_0\leq 0.35]$ is also the region where the ratchet current $\bar v$ is 
appreciable. The P\'{e}clet number , $P_e$, as defined earlier, are also calculated as a 
function of $F_0$. They are plotted in the inset of Fig.2. It is clear from the figure 
that in the same region, $P_e$ is also much larger than 2. This indicates that in the 
region $[0.15\leq F_0\leq 0.35]$ the particle motion is highly diffusive but 
concomitantly it is greatly coherent too. This is also indicated by the observation that 
even though the position dispersions (fluctuations) are large the relative fluctuations 
of position in this region are considerably low ($<1$). As indicated by the result in 
the adiabatic case (Fig. 3 \cite{wanda}) this range of $F_0$ of coherent motion is 
expected to shift as the value of $\gamma_0$ is changed.

Though our system is different from that of Machura, et. al. \cite{machura}, at this 
point it would be interesting to make a comparison with their result. They observe that 
for their low temperature case $D_0=0.01$ in the vicinity of $a\approx0.6$ the velocity 
fluctuation underwent a rapid change (Fig.1a of \cite{machura}). To translate this to 
our case\cite{comp} $a\approx0.6$ is equivalent to $F_0\approx 0.2$ and given their 
potential barrier being just about half of the value in our case one should expect the 
peaking of velocity dispersion to occur below $F_0=0.4$. Taking into consideration of 
our temperature ($T=0.4$) being 40 times 0.01 the phenomena should occur much below 
$F_0=0.4$. In this sense the region $[0.15\leq F_0\leq 0.35]$ seems quite reasonable. 
Also, $\bar v$ of Fig.3a of Ref. \cite{machura} at $D_0=0.4$ make a good comparison 
with Fig.1 in our case. However, as mentioned earlier the two systems are quite 
different in basics to have an exact comparison.

\begin{figure}
\begin{center}
\epsfig{file=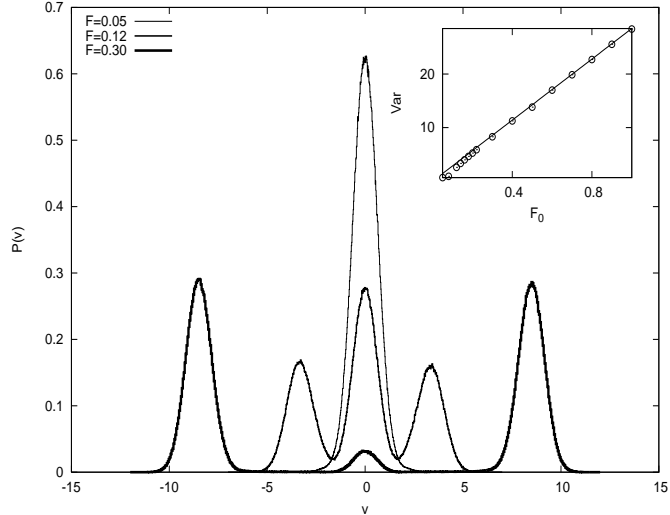, angle=-90, width=9cm,totalheight=7cm}

\caption{Plot of velocity distribution $P(v)$ for three values of driving amplitudes 
$F_0=0.05$, 0.12 and 0.30. The figure in inset shows the variance of velocities as a 
function of $F_0$ fitted with a straight line to show the linear growth of variance at
large $F_0$.
\label{fig3}}

\end{center}
\end{figure}

The velocity distribution $P(v)$ also shows interesting behaviour. In Fig.3, we plot 
$P(v)$ for three values of $F_0$. A sharp peak which is almost indistinguishable from a 
Gaussian centred at $v=0$ for small $F_0=0.05$ gets split up into three peaks for 
$F_0=0.12$, and similarly for $F_0=0.30$, with the central peak, gradually diminishing. 
This shows a behaviour, including the nearly linear growth of the variance with $F_0$ 
(inset, Fig.3), quite similar to what has been reported earlier in a different
system\cite{machura}. There is, however, one difference. The side peaks of $P(v)$ in our 
calculation have origin in the running states of the particle. It is, perhaps, due to 
the square-wave drive, instead of sinusoidal drive, that for as low amplitude as
$F_0=0.3$ we get three disjoint velocity bands and at $F_0=0.6$ we get just two bands, 
the central band being almost unpopulated. The three peaks, for example for $F_0=0.3$, 
could be fitted to a combination of three Guassians. With a cursory look, the left and 
right Gaussians barely show much difference. However,
\begin{equation}
\langle v \rangle = \int_{-\infty}^{\infty} v P(v) dv
\end{equation}
gives approximately the same value as $\bar v$, and $\langle v \rangle (F_0)$ showing 
exactly the same nature as $\bar v(F_0)$ (Fig.4).

\begin{figure}
\begin{center}
\epsfig{file=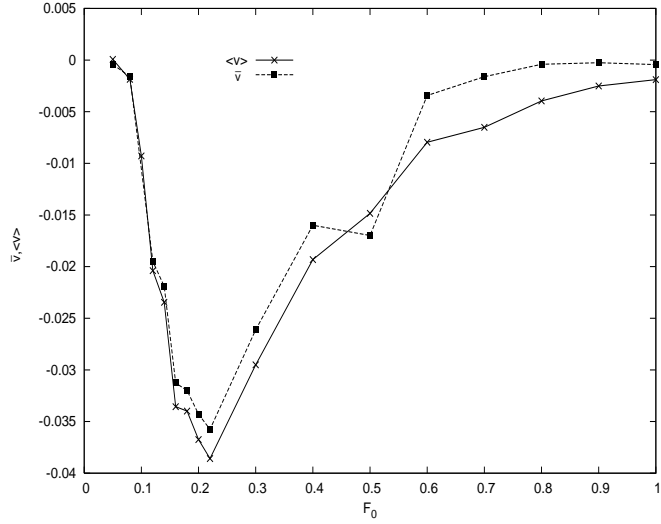, angle=-90, width=9cm,totalheight=7cm}

\caption{Shows the variation of the steady state mean velocity  $\bar v$, Eq. (2.3) and 
$\langle v \rangle$, Eq. (3.2) for the same parameter values as in Fig.2.
\label{fig4}}
\end{center}
\end{figure}

\begin{figure}
\begin{center}
\epsfig{file=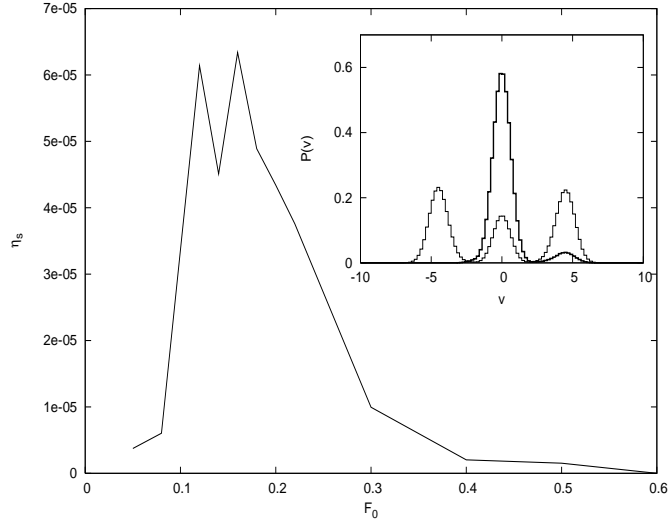, angle=-90, width=9cm,totalheight=7cm}

\caption{Shows Stokes efficiency, $\eta_{S}$ as a function of $F_0$ for the same
paramater values as in Fig.2. The inset shows the difference in the velocity 
distribution for symmetric (three peaks) and asymmetric drive for the same value of
$F_0=0.16$ and $\tau =2000$ and $\alpha=0.2$.
\label{fig5}}

\end{center}
\end{figure} 

\subsection{The efficiency of ratchet performance}
From the velocity distribution $P(v)$ we calculate the Stokes efficiency, $\eta_{S}$, 
defined as\cite{machura},
\begin{equation}
\eta_{S}=\frac{\langle v \rangle ^2}{|\langle v^2 \rangle - T|},
\end{equation}
as a function of $F_0$. Fig.5 shows that $\eta_{S}$ is larger in the same range
of $F_0$ where it shows larger $\bar v$. The peak of $\eta_{S}$, however, does not
occur at the same position as the peak of $\bar v$. It is, however, to be noted that the
plotted figure is calculated from averages over a small number ($\sim 20$) of ensembles
because it is computationally quite expensive to obtain results for the steady state
(maximum $t=10^7$) and hence not feasible to obtain averaging over a larger number of
ensembles. Though the qualitative behaviour is encouraging the efficiencies are 
small $\sim 10^{-5}$. In the adiabatic drive case (part I, Ref.\cite{wanda1}) we have 
found that Stokes 
efficiency depends on various parameter values: $\gamma_0$, $T$, etc. The efficiency 
shown here is for a small $\gamma_0=0.035$, $T_\Omega=1000$, and $T=0.4$ where the 
current is also very low. The efficiency of 
this symmetrically driven system can, however, be improved to a good extent by an 
optimal choice of these parameters. 

An inertial ratchet driven by a zero mean asymmetric drive can, however, give a highly
efficient performance compared to the symmetrically driven ratchet. For example, when
the system is driven by a field
\[ \begin{array}{llll}
  F(t)&=&\pm F_0,& ( n\tau\leq t < (n+\alpha)\tau),\\
 &=&\mp\frac{\alpha}{(1-\alpha)}F_0,& ( (n+\alpha)\tau \leq t < (n+1)\tau),\\
\end{array}\] \\
with $\alpha = 0.2$ gives an efficiency of $3.8\times 10^{-2}$ compared to 
$6.2\times 10^{-5}$ in the symmetric-drive ($\alpha=0$) case with $F_0=0.16$ and 
$\tau=2000$. This is 
made possible because in the symmetric drive case the particles move on either direction 
with almost equal probability whereas in the asymmetric drive case the particle motion in 
one direction is practically blocked, as is evident from the corresponding velocity
distributions shown in the inset of Fig.5. The contribution of the system inhomogeneity 
for this improved performance is, however, quite insignificant.

\subsection{The transient-state dispersions and the ratchet current} 
When a constant force $F$ is applied to the system it shows dispersionless behaviour:
$\langle (\Delta x(t))^2 \rangle$ does not change with time in the intermediate time 
scales, roughly [$10^3<t<10^5$], for around [$0.12<F<0.7$] at $T=0.4$ for 
$\gamma_0=0.035$. The result of dispersionless behaviour had originally been shown and 
explained\cite{linde} beautifully for constant friction $\gamma$ case: the position
distribution moves undistorted at constant velocity $v=\frac{F}{\gamma}$ or 
equivalently, velocity distribution remains undistorted centerd at 
$v=\frac{F}{\gamma}$. The interval [$t_1<t<t_2$] of time $t$ during which the system 
shows this remarkable intermediate-time behaviour depends on the tilt force $F$, as 
should also on other parameters. $t_1$ is roughly of the order of but much larger than 
the Kramers passage time corresponding to the lower of the potential barriers on either 
side of a well. The transient-time dispersionless particle-motion behaviour is sensitive 
to initial conditions. In the following we specifically begin from the bottom of the 
well at $x=\frac{\pi}{2}$ with particle velocities appropriate to the Boltzmann 
distribution at temperature $T=0.4$.
\begin{figure}
\begin{center}
\epsfig{file=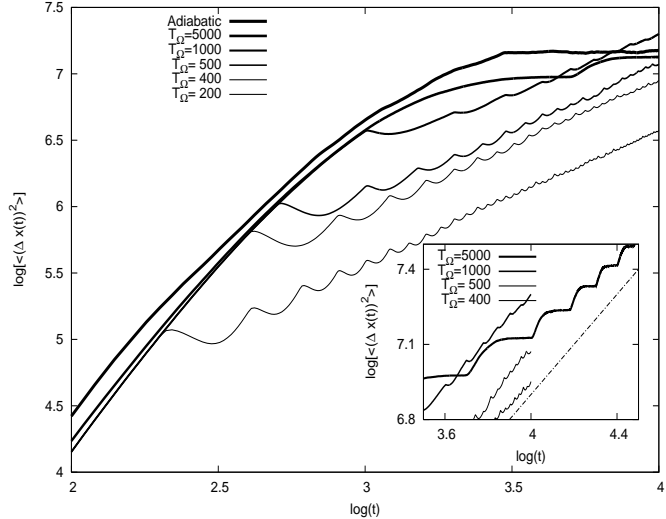, angle=-90, width=9cm,totalheight=7cm}

\caption{The plot of position dispersions $\langle (\Delta x(t))^2 \rangle$ versus time 
$t$ (in logarithmic scale) for different values of ($T_{\Omega}$) of forcing with 
$F_0=0.2$. The inset shows the clipped part of the plot at larger time.
\label{fig6}}

\end{center}
\end{figure} 

When the inhomogeneous system is driven periodically by a sqaure-wave forcing of 
amplitude $F_0$, the dispersionless coherent nature of average motion gets interrupted 
depending on the period $T_\Omega$ of the forcing [Fig.6]. When $t_1<T_{\Omega}<t_2$, at 
$t=T_{\Omega}$ the dispersion gets a jerk and shoots up only to get flattened again to 
an another bout of dispersionless regime. This regime too gets a similar jolt after 
another $T_{\Omega}$ and the process continues for a large number of periods. When the 
direction of the applied force is changed the 'forward moving' particles are forced to 
halt momentarily to begin moving in the new direction of the force afresh. While in the 
state of halt particles are more likely to find themselves closer to the bottom of some 
well and thus the system gets initialised as in the beginning. The system finds itself 
in similar situation again and again periodically with each change of force direction 
and continues with its unfinished dispersionless sojourn for a large number of periods 
with remarkable robustness [Inset of Fig.6]. However, when $T_{\Omega}<t_1$ the system 
never gets a chance to experience its dispersionless journey because only a fraction of 
the particles get the opportunity to acquire the required constant average 
velocity\cite{linde} of $\frac{F_0}{\gamma_0}$ and the rest keep lagging behind even by 
the end of constant force duration $T_{\Omega}$. Instead, as soon as the direction of 
the force $F$ is reversed, after $T_{\Omega}$ the dispersion dips after a brief climb 
up, as the particles get herded together briefly before getting dispersed further in the 
reversed direction of $F_0$. This can be seen very clearly in the time evolution of the 
position probability distribution profile $P(x,t)$. The front of the $P(x,t)$ moves with 
velocity $\frac{F_0}{\gamma_0}$ while the rest lag behind it moving at a slower speed 
but trying to catch up with the front throughout $T_{\Omega}$. This process of dispersion
dipping (after a small contunuing rise) and rising to a higher value after each 
$T_{\Omega}$ is repeated for several tens of periods [Fig.6].
    
The intermediate-time dispersionless motion is not an exclusive characteristic feature 
of inhomogeneous systems. It is a characteristic feature of inertial washboard potential 
system\cite{linde}. However, its study in the inertial inhomogeneous system provides a 
convincing explanation of the variation of ratchet current as a function of $T_{\Omega}$ 
[inset of Fig.1] and helps in finding a criterion to improve the performance of the 
ratchet.

\begin{figure}
\begin{center}
\epsfig{file=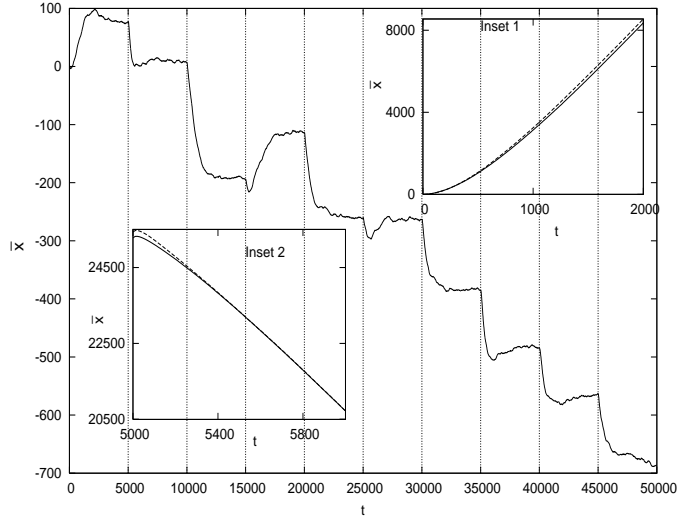, angle=-90, width=9cm,totalheight=7cm}

\caption{The average displacement of particles as a function of time, driven by equal 
number of $\pm F(t)$ profiles (or equal number of odd and even numbered trajectories) 
for $F_0=0.2$, and $T_{\Omega}=5000$. The insets highlight the
contributions to the mean displacement of odd and even numbered trajectories separately,
leading to the main figure. The mean displacements for the even numbered trajectories
are shown with a reversed sign.
\label{fig7}}

\end{center}
\end{figure}
 In fig.7 the average displacement of particles as a function of time when driven by 
equal number of $\pm F(t)$ profiles is presented for $F_0=0.2$, and $T_{\Omega}=5000$. 
This case corresponds to the repeated 
dispersionless motion shown in Fig.6. Fig.7 clearly shows that during the dispersionless 
motion the average dispacement of particles effectively remains constant. In other words,
during the period of dispersionless motion the particles move equally in the left as 
well as in the right direction thereby contributing nothing to the ratchet current: 
while in the dispersionless motion the particles fail to see the frictional 
inhomogeneity of the system. All the change in the average displacement and hence all 
the contribution to the ratchet current comes during the dispersive period of motion. 
This is shown in the inset of Fig.7 where for clarity the mean particle positions for 
$F(t)$ beginning with $F(0)=-F_0$ (even numbered trajectories) are shown as a function 
of time with their sign reversed. The mean particle displacements for odd and even 
numbered trajectories differ only during the interval just after the reversal of $F_0$ 
and before the dispersionless regime begins and the two lines of mean positions 
(insets of Fig.7) run parallel during the dispersionless regime. This clearly indicates 
that in order to get a larger current an optimum choice of $T_{\Omega}$ needs to be made 
which, naturally, avoids the dispersionless regime but is not too small in order to 
allow the particles to leave their potential wells. This conclusion is well supported by 
the inset of Fig.1.

\begin{figure}
\begin{center}
\epsfig{file=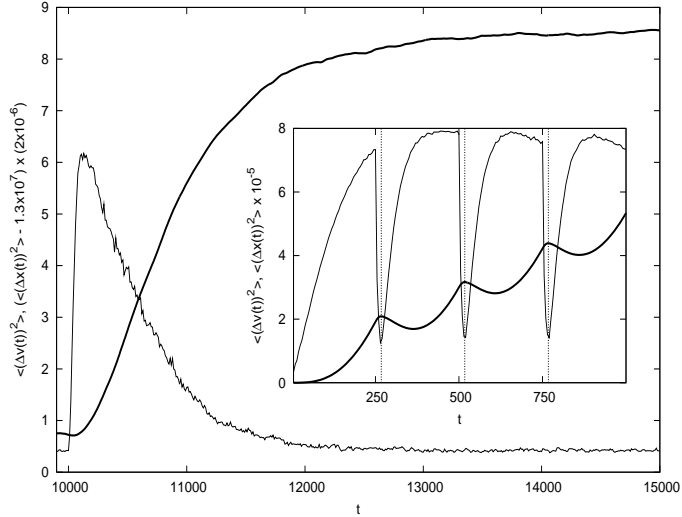, angle=-90, width=9cm,totalheight=7cm}

\caption{Illustration of velocity dispersions $\langle (\Delta v(t))^2 \rangle$ (thin 
line) and position dispersions $\langle (\Delta x(t))^2 \rangle$ (bold line) during a 
time interval for $T_{\Omega}=5000$. The inset shows the corresponding plots for square 
drive forcing with smaller $T_{\Omega}$=250 with no dispersionless regime. The thin vertical lines roughly indicate the positions of the extrema of the dispersion curves. 
\label{fig8}}

\end{center}
\end{figure}

The velocity dispersions and position dispersions together show interesting behaviour.
Fig.8 shows that during the dispersionless regime when the position dispersion is 
constant and maximum the velocity dispersion is also constant but it has a minimum
value. This minimum constant value is repeated in all the $nT_{\Omega}, n=1,2,...$ 
intervals whereas the value of the constant position dispersion increases in every 
successive $nT_{\Omega}$ interval as shown in Fig.6. In the dispersive regimes the 
velocity dispersions are squeezed to very sharp troughs exactly where the position 
dispersions show sharp peaking. In the inset of Fig.8 these dispersions are shown for 
$T_{\Omega}=250$. The onward rush of the particles do not halt immediately after the 
direction of $F_0$ is changed at $nT_{\Omega}$ but it continues for a very short time 
giving a small increase in the spread of $P(x)$. Then a majority of particles stop, 
giving a sharp peak in the $P(v)$ at $v=0$ reducing its spread drastically. At that 
moment the product of position and velocity distribution spread becomes a minimum. 
The reverse journey thereafter increases the spread of $P(v)$ but there is a slow 
squeezing of $P(x)$ before it begins to spread again. The maximum $P(x)$ squeezing,
however, does not exactly coincide with the largest of the broad $P(v)$ but it is at a 
rather closer range. In this case too the minimum velocity dispersion remains constant 
for all $nT_{\Omega}$. But the wings of $P(x)$, though thin, keep spreading with time 
giving an average increase of dispersion as time increases. However, most of the 
particles remain confined roughly to a region 
[$-\frac{|F_0|}{\gamma_0}T_{\Omega}<x<+\frac{|F_0|}{\gamma_0}T_{\Omega}$] for a long time.

\section{Discussion and conclusion} 
The ratchet effect, in this work, is brought about just by the phase lag $\phi$ between 
the potential and the friction of the medium, without having to have an external bias. 
This is seemingly a weak cause to generate unidirectional current. The Figs.1 through 5 
refer to a square-wave forcing with $T_{\Omega}=1000$. The choice of this $T_{\Omega}$ 
clearly avoids the dispersionless regime. Yet, this is not the optimum value of 
$T_{\Omega}$. It should have been around 500 in order to get the largest possible 
ratchet current. This choice would have definitely enhanced the efficiency of operation. 
The same can also be said about other parameters, such as $T$, and $\phi$ for 
$\gamma_0=0.035$. However, with the help of these figures we have been able to exhibit 
the qualitative trends shown by the ratchet.

In the inset of Fig.6 we have drawn a straight line with slope 1 as a guide to show that
ultimately the curves should achieve that average slope at large times for various 
$T_{\Omega}$ values of drives. Even though the average slope of the curves have not yet 
reached the diffusive slope of one the small $T_{\Omega}$ curves are slowly approaching 
that value. One can,therefore, safely infer that in the steady state situation the 
effective diffusion constant should increase monotonically with $T_{\Omega}$ for small 
$T_{\Omega}$. The frequency of drive or equivalently $T_{\Omega}$, thus, plays important 
role about how the particles diffuse out of their wells. For example, The population of 
the initial well depletes with time exponentially, $N(t)=N(0)e^{-bt}$, with $b=0.0023$
for $T_{\Omega}=250$ and $b=0.002$ for $T_{\Omega}=500$, for $\gamma_0=0.035$ at 
$T=0.4$ that we have studied. By the time the well gets effectively exhausted the 
first particles would have moved farther than a thousand of potential wells. Of course,
this first well itself (as all others) keeps getting repopulated all the time.

The dispersive behaviour for drives with $T_{\Omega}>t_2$ is difficult to study because
it takes a very large computer time to arrive at a concrete result. However, the 
indications are there that for these large $T_{\Omega}$ also, the system will show 
repeated dispersionless regimes, though somewhat enfeebled because the process of 
diffusion will dominate at these large times.

To conclude, the study suggests an interesting method of obtaining ratchet current
in inertial noisy systems by exploiting the frictional inhomogeneity of the medium. It
also exhibits clearly that the ratchet current, in this system, is contributed by
dispersive conditions and not by coherent movements of particles. 

\section*{ACKNOWLEDGEMENT}

MCM acknowledges BRNS, Department of Atomic Energy, Govt. of India, for partial
financial support and thanks A.M. Jayannavar for discussion, and Abdus Salam ICTP,
Trieste, Italy for providing an opportunity to visit, under the associateship scheme, 
where a part of the work was completed and the paper was written.


\begin{thebibliography}{34}
\bibitem{comp} The parameter scaling is different from that of Machura, et. al., where 
\cite{machura} the forcing amplitude $a$ is equivalent to $\pi$ times the amplitude 
$ F_0$ in our case. The total potential barrier height (at $a=0$) is $\sim 1$ , whereas 
in this work it is 2. However, the noise strength $D_0$ in Ref. \cite{machura} is same
as the temperature $T$ used here.
\end{thebibliography}
\end{document}